\pdfoutput=1
\documentclass[twocolumn,nofootinbib,preprintnumbers,amsmath,amssymb]{revtex4}
\newcommand{\ba}{\begin{eqnarray}}
\newcommand{\ea}{\end{eqnarray}}

\usepackage[pdftex]{graphicx}
\usepackage{epsfig}
\usepackage{amssymb}

\begin{document}

%\preprint{MCTP-11-} 

\title{Mixed Wino-Axion Dark Matter in string/$M$ theory and \\ the 130 GeV $\gamma$-line ``Signal"}
\author{Bobby Samir Acharya$^{1,2}$}
\author{Gordon Kane$^{3}$}
\author{Piyush Kumar$^{4}$}
\author{Ran Lu$^3$}
\author{Bob Zheng$^3$}
\affiliation{$^1$Department of Physics, King's College London, UK\\
$^2$International Centre for Theoretical Physics, Trieste,  Italy\\
$^3$Michigan Center for Theoretical Physics, Ann Arbor, MI 48109 USA \\
$^4$Department of Physics, Columbia University, New York, NY 10027 USA}
%\date{\today}
% It is always \today, today

\begin{abstract}
String/$M$ theory compactifications with low energy supersymmetry tend to predict that dark matter has two components: axions and WIMPs \cite{Acharya:2010zx,1204.2795}. In accord with this, we show that the tentative 130 GeV $\gamma$-line signal reported in \cite{1204.2797} can be interpreted as arising from the 
annihilation of 145 GeV mass, Wino-like WIMPs into a $Z$-boson and a photon. In this context, the signal implies a second component of dark matter which we interpret
as being composed of axions - the relative Wino/Axion abundances being approximately equal.
Further predictions are implied: signals in both diffuse and monochromatic photons from dwarf spheroidal galaxies; monochromatic photons with energy 145 GeV; for the LHC, the Higgs boson mass has been predicted in this framework    
\cite{1112.1059}, and the current Higgs limits provide interesting constraints on the mass of the Gluino.   
\end{abstract}

%\pacs{11.25.Mj 11.25.Wx 11.25.Yb 12.10.-g 12.60.Jv 14.80.Ly}% PACS, the Physics and Astronomy

\maketitle

%\tableofcontents

%\newpage

\section{Introduction}\label{signal}

Although the existence of Dark Matter (DM) is a well established fact, both its origin and composition remain a deep mystery. 
Arguably the two most theoretically well-motivated DM candidates are axions \cite{Wilczek:1977pj} and weakly interacting massive particles (WIMPs). WIMP candidates 
for DM often exist within the context of supersymmetry. Both WIMPs and axions are present in solutions of string/$M$ theory with low energy supersymmetry breaking and
it has been argued that both should form a significant component of DM \cite{Acharya:2010zx,1204.2795}. In fact, considerations of cosmological moduli physics imply that the WIMP
component of dark matter is Wino-like\footnote{The Wino is the spin 1/2 superpartner of
the $W^0$ gauge boson in the Standard Model.} \cite{arXiv:0801.0478, Acharya:2008bk, Acharya:2009zt}.
%As the name suggests, in addition to gravity WIMPs couple to each other and to SM particles via weak interactions. This has led to a considerable amount of effort in the community trying to detect the existence of WIMPs via their scattering against nucleons (``Direct" detection), or via their self-annihilation in the Universe leading to cosmic-ray fluxes (``Indirect" detection). Data from these experiments and observations have already provided stringent limits on many models, for instance ruling out large regions of the MSSM parameter space.

WIMPs can be ``indirectly" detected by observing the cosmic ray fluxes arising from their annihilation. Among the cosmic rays, $\gamma$-rays are special because they propagate essentially unperturbed and retain spatial information about their sources. In addition to providing a diffuse $\gamma$-ray signal, DM annihilations also give rise to monochromatic $\gamma$ `lines', whereas the astrophysical background tends to be a `featureless' continuum. 
Thus, clear detection of a monochromatic $\gamma$-ray signal is thought to be `smoking-gun' evidence for DM \cite{Bergstrom:1988fp}. 

In light of this, the recent claim in \cite{1204.2797} of a tentative $\gamma$-ray line signal in the Fermi satellite data at an energy $E_{\gamma}\approx 130$ GeV  - arising from the Galactic Center (GC) with a local (global) significance of 4.6$\,\sigma\,(3.3\,\sigma)$ - is extremely tantalizing. An independent analysis by \cite{Tempel:2012ey} corroborates the existence of such a signal. Further studies of these claims are eagerly awaited.
%Admittedly, the signal is based on the observation of around 50 photons which is not that large.  Also, all systematic effects may have not been taken into account. However, the reported claim is based on analysis of 43 months of data and a careful selection of target regions close to the GC. Furthermore, the statistical method was examined with a bootstrap analysis to exclude obvious systematic uncertainties. Finally, the fact that the signal peaks around the GC makes a purely instrumental uncertainty a very unlikely origin of the reported signal.
The Fermi-LAT collaboration has not officially reported a $\gamma$-line signal; however, they have recently set limits on the DM annihilation cross-section to photons by searching for $\gamma$-lines from essentially the entire Milky Way (excluding the Galactic Plane) \cite{Ackermann:2012qk}, thus complementing the target regions close to the GC studied in \cite{1204.2797}. In any
case, confirmation (or not) of the signal could come from several experiments in the near
future e.g. the observation of the $\gamma$-line signal  from the GC and dwarf spheroidal galaxies, as well as the observation of excess diffuse $\gamma$-rays from the dwarf galaxies, by Fermi-LAT and AMS-2 . A second $\gamma$-line signal at nearby energies is predicted by many models and this should also be observed by the two experiments.

In this paper, we will assume that this tentative $\gamma$-line signal does indeed arise from the annihilation of a WIMP component of DM, and we interpret the broad consequences
for physics beyond the Standard Model, focusing in particular on low energy supersymmetry models which arise from string/$M$ theory. We will demonstrate that, consistent with the generic predictions \cite{1204.2795}, not only can the signal be explained well with Wino-like WIMPs\footnote{The Wino is the spin 1/2 superpartner of
the $W^0$ gauge boson in the Standard Model.}, it also {\it implies a second component of DM}, which we interpret as being composed of axions. 
%We emphasize that the generic prediction that DM consists of non-negligible fractions of WIMPs and axions was made \emph{before} the appearance of the data \cite{Acharya:2010zx, 1204.2795}, although  the precise ratio of the WIMP and axion abundances was not computable from first principles. 
The data determines the LSP mass and also the annihilation rate to photons (with uncertainties), so we can deduce a number of consequences, the estimate of the ratio of the WIMP and axion abundances being one of them. 

In section \ref{implications}, the broad implications of the signal for BSM physics from a low energy point of view are discussed. Then, 
in section \ref{string}, 
we summarize a top-down approach which studies generic properties of realistic string vacua (see \cite{1204.2795} for a broad review), emphasizing how it leads to
the prediction that DM is composed of both WIMPs and axions. 
In section \ref{signal-string}, we demonstrate that the $\gamma$-line signal is consistent with Wino-like WIMPs and we determine the ratio of the axion and WIMP abundances.  
In section \ref{associated}, we study the constraints and associated WIMP signals of the framework. In particular, we show that the current limits from searches for $\gamma$-lines from the Milky Way \cite{Ackermann:2012qk} as well as searches for excess diffuse $\gamma$-rays  from dwarf spheroidal galaxies \cite{Garde:2011hd} also suggest a non-negligible contribution to DM which is \emph{not} WIMPs. Finally, in section \ref{falsifiable}, we make correlated predictions for particle physics. For example, in the particular framework which gives a Higgs mass prediction \cite{1112.1059} compatible with the recent hints at the LHC, the Gluino and Bino masses are predicted as functions of the gravitino mass $m_{3/2}$.

\section{Required Cross-section and Implications}\label{implications}

The most interesting feature of a $\gamma$-line signal is that the energy of the line $E_{\gamma}$ is very simply related to the mass of the WIMP due to the WIMPs in the halo being almost at rest. Since the signal is at $E_{\gamma} \approx 130$ GeV, this means that if the signal is interpreted as coming from WIMP annihilation to $\gamma\,\gamma$, then the WIMP mass $m_{\chi} \approx 130$ GeV. If the signal is instead interpreted as arising from $Z\,\gamma$ final states, this would imply $m_{\chi} \approx 145$ GeV. 
%due to the phase space factor $(1-\frac{m_Z^2}{4\,m_{\chi^2}})$.  

There is considerable uncertainty in extracting the best-fit annihilation cross-sections into $\gamma\gamma$ or $Z\gamma$, even if the signal is correct. The analysis of \cite{1204.2797} gives the results: 
{\setlength{\arraycolsep}{1pt}\ba\label{reqd}\langle \sigma v\rangle^{reqd}_{\chi\chi\rightarrow \gamma\,\gamma} &\approx& 1.27 \pm 0.32^{+0.18}_{-0.28}\times10^{-27}{\rm cm^3/s}:\;\;{\rm Einasto}\nonumber\\
&\approx& 2.27\pm 0.57^{+0.32}_{-0.51}\times10^{-27}\,{\rm cm^3/s}:\,{\rm NFW}\ea} for the two DM profiles with $\rho_{DM}^{sun}$ being normalized to $0.4\,{\rm GeV/cm^3}$. As is known, the local DM density at the sun's position has an ${\cal O}(1)$ uncertainty. The presence of DM substructures very close to the GC can also lead to an increased $\gamma$-ray flux. Furthermore, \cite{Tempel:2012ey} claims that scanning the Galaxy in a way such as to \emph{maximize} the 130 GeV signal gives rise to a larger flux and increases the required cross-section by a factor of few.  In this work, for concreteness we study the implications of the more conservative analysis by \cite{1204.2797}, in which the target regions for study were chosen beforehand and the existence of the signal was then determined. Nevertheless, it is important to remember that the results of this paper are only valid within the set of caveats mentioned above. 

The cross-section in (\ref{reqd}) is large compared to naive expectations for the following reason. The standard assumption which is made about WIMP DM is that it is a thermal relic. With this assumption, the WIMP annihilation cross-section at thermal freezeout is $\langle \sigma\,v^{tot}_{\chi\chi}\rangle = \langle \sigma\,v^{therm}_{\chi\chi}\rangle \approx 3\times 10^{-26}\,{\rm cm^3/s}$. However,
WIMPs typically annihilate to photons only via loop effects. Hence, the cross-section is suppressed by a loop factor: 
\ba  \langle \sigma v\rangle_{\chi\chi\rightarrow \gamma\,\gamma}  \lesssim \frac{1}{16\pi^2}\,{\langle \sigma\,v^{therm}_{\chi\chi}}\rangle \approx 1.9\times 10^{-28}\,{\rm cm^3/s}, \ea which is at least an order of magnitude {\it smaller} than the reported signal. This is true for the minimal supersymmetric standard model
(MSSM) and many other models. On the other hand one could reconcile a thermal cross-section with significant annihilation into photons by considering either models in which the annihilation to photons is somehow enhanced \cite{Dudas:2012pb}, or in which there exist very light states giving rise to a Sommerfeld enhancement of the cross-section \cite{Tempel:2012ey}. In this work, we study a different and rather appealing mechanism which we will show is compatible with the tentative signal as well as all other contraints, and moreover, works for simple and well motivated models such as the MSSM: this is the non-thermal WIMP `miracle' \cite{Acharya:2009zt} and is reviewed below. 

\subsection{The non-thermal WIMP `Miracle'}\label{nonthermal}

Compactified string/$M$ theory generically gives rise to moduli fields in the effective low energy description of physics. These are the low energy manifestations of the extra dimensions present in string/$M$ theory and are necessarily present as long as the supergravity approximation is valid. Moduli fields couple fairly universally to matter with interactions suppressed by a large scale, such as the GUT or Planck scale. They generically will dominate the energy density of the Universe after inflation but must decay before big-bang nucleosynthesis (BBN). When they decay, they not only dilute the density of any thermal relics by many orders of magnitude, but they also produce WIMPs as decay products.

This gives rise to a WIMP number density of order :
\ba \label{ntwm} n_{\chi} \sim \frac{\Gamma_{X}}{\langle\sigma\,v\rangle^{tot}_{\chi\chi}} \sim \frac{H(T_R)}{\langle\sigma\,v\rangle^{tot}_{\chi\chi}}\ea   where $T_R$ is the reheat temperature generated when the modulus $X$ decays, $\Gamma_X$ the modulus decay width and $H(T)$ the Hubble scale at temperature $T$. 
By contrast, in the thermal case, $n_{\chi}^{thermal}\sim \frac{H(T_F)}{\langle\sigma\,v\rangle^{tot}_{\chi\chi}}$, where $T_F$ is the WIMP freezeout temperature. This implies that to obtain a roughly correct abundance in the non-thermal case, $\langle\sigma\,v\rangle^{tot}_{\chi\chi}$ has to be larger compared to that for the thermal case by a factor of $\frac{T_F}{T_R}$. Furthermore, with $\langle\sigma\,v\rangle^{tot}_{\chi\chi}$ larger by a factor of $\frac{T_F}{T_R}$ relative to the thermal case, the one-loop suppressed $Z\,\gamma$ ($\gamma\,\gamma$) channels can naturally have a cross-section consistent with the tentative Fermi-LAT signal! Finally, within the framework of supersymmetry there naturally exist WIMP LSP candidates like the Wino or the Higgsino with annihilation cross-sections precisely in the required range. Since Winos have a larger annihilation cross-section than Higgsinos for the same mass, a Wino-like LSP with a small Higgsino component is favored\footnote{For simplicity, we do not consider models beyond the MSSM.}.

In the next section, we describe an approach which satisfies all the conditions required for the ``non-thermal WIMP miracle" to work, and which also gives rise to Wino-like LSPs in a large region of parameter space. Henceforth, we will study this case in detail.

\section{Generic DM Predictions in String/$M$ theory}\label{string}

In this section, we briefly summarize a top-down framework which aims to study generic properties of  large classes of realistic solutions of string/$M$ theory (with rather mild assumptions). In particular, by ``realistic" we mean vacua which have stabilized moduli with supersymmetry breaking at a low scale, a compactification scale $M_{KK}$ around the traditional GUT scale, and which satisfy all phenomenological and cosmological constraints.  For a broad review of this approach see \cite{1204.2795}. 

The key to making these generic predictions lies in the physics of the moduli fields whose vacuum expectation values parametrize the size and shape of extra dimensions, and appear in the low energy effective supergravity theory in four dimensions. 
In supergravity, the gravitino mass $m_{3/2}$ generically sets the scale for all scalar particles, including both the moduli and MSSM scalars. This is borne out by explicit string/$M$ theory examples. There are two interesting exceptions - the axions and the lightest Higgs boson. The scalar trilinear couplings are also ${\cal O}(1)\,m_{3/2}$. The actual value of $m_{3/2}$ is set by cosmological arguments. Requiring that the moduli decay before BBN leads to\footnote{Much larger values of $m_{3/2}$ are disfavored by
considerations of axion physics \cite{Acharya:2010zx, 1204.2795} and the Higgs mass \cite{1112.1059}.} $m_{3/2} \gtrsim$ 30 TeV.  This implies that most of the history of the pre-BBN is matter dominated by moduli and not
radiation dominated as is often assumed.

The axion fields $a_i$, which are pseudoscalar partners of the moduli, do not get masses of ${\cal O}(m_{3/2})$ because of shift symmetries $a_i\rightarrow a_i + c_i$ which originate from gauge symmetries in higher dimensions. Non-perturbative effects will stabilize these axions with masses exponentially suppressed relative to $m_{3/2}$. Since there are typically large numbers of axions, their masses are distributed roughly linearly on a logarithmic scale, as suggested in \cite{Arvanitaki:2009fg}. One of these light axions could naturally be the QCD axion, solving the strong CP problem together with moduli and axion stabilization \cite{Acharya:2010zx}. Because of their tiny masses, many of these axions start oscillating when the moduli are dominating the energy density of the Universe. Hence, when the moduli decay releasing a large amount of entropy, this dilutes the energy density of axions significantly. 
This gives an upper bound on the axion decay constant $f_a$ of order $10^{15}$ GeV, close to the natural value obtained in these compactifications of around $M_{GUT} \sim 10^{16}$ GeV \cite{Acharya:2010zx}.  Thus, with a small ($\sim 1\%$) tuning an ${\cal O}(1)$ abundance of DM is predicted to be in the form of axions. Note that assuming a radiation dominated pre-BBN Universe leads to an upper bound $f_a < 10^{12}$ GeV with ${\cal O}(1)$ misalignment angle, which is typically inconsistent with Grand Unification.

In supersymmetric models with a sufficiently conserved stabilizing symmetry (like $R$-parity), the LSPs will also be present in the early Universe as WIMP DM candidates. Even though moduli decay essentially wipes out their thermal abundance, they are regenerated by the moduli decay themselves.  Although scalar masses are generically ${\cal O}(m_{3/2})$, gaugino masses need not be, either due to symmetries or dynamics. There exist string/$M$ theory solutions where the gaugino masses are naturally suppressed relative to $m_{3/2}$ (see \cite{Acharya:2006ia}, \cite{hep-ph/0511162});  the LSP can then be Wino-like (possibly with a very small Higgsino component) in a large region of parameter space, and non-thermal production by moduli decay can naturally provide an ${\cal O}(1)$ abundance of DM, via the so called ``non-thermal WIMP miracle", which was shown in \cite{Acharya:2008bk} and is discussed in section \ref{nonthermal}. 

Thus, we see that the framework generically predicts two non-negligible sources of DM - axions and WIMP LSPs. The precise ratio of abundances of the two components cannot be determined from the theory yet. \emph{This is where the Fermi $\gamma$-line signal comes in}. It provides a rather accurate determination of the LSP mass. We will see in the following sections that modulo astrophysical uncertainties, the tentative signal determines the relative fraction of dark matter in the form of WIMPs and axions.  Specific predictions for other observables also arise, most importantly the mass of the Gluino. Mixed WIMP/axion dark matter has also been considered recently in \cite{Baer:2011hx}.

Before moving on, we note that when the matter and gauge spectrum below the GUT scale is precisely that of the MSSM, the mass of the lightest Higgs boson can be accurately computed in this framework \cite{1112.1059} giving results in good agreement with the recent hints from the ATLAS and CMS experiments 
\cite{ATLAS-CMS-Higgs}. This provides a compelling reason to interpret the tentative DM signal in this context.

\section{Interpreting the signal in string/$M$ theory.}\label{signal-string}

The $\gamma$-ray flux for Wino-like WIMPs has a bigger contribution from $Z\,\gamma$ rather than $\gamma\,\gamma$ final states. The flux is given by the line of sight integral of the square of the DM density profile, normalized by the cross-section: \ba\label{flux} \frac{d\,\Phi_{\gamma}}{dE\,d\Omega} (\xi) = \frac{\langle \sigma\,v\rangle_{\chi\chi\rightarrow Z\gamma}}{8\pi\,m_{\chi}^2}\,2\delta(E-E_{\gamma})\,\int_{l.o.s.}\,ds\,\rho_{\chi}^2\,(r)\ea where $\xi$ is the angle to the GC, $m_{\chi}$ is the WIMP mass, $\langle \sigma\,v\rangle_{\chi\chi\rightarrow Z\,\gamma}$ is the partial annihilation cross-section of the WIMPs to $Z\,\gamma$, $E_{\gamma}=m_{\chi}\left(1-\frac{m_Z^2}{4\,m_{\chi}^2}\right)$ is the $\gamma$-line energy and $\rho_{\chi}(r)$ is the DM profile as a function of the Galactocentric distance $r$.

The tentative $\gamma$-line signal when interpreted in terms of $\gamma\gamma$ final states yields an effective cross-section of $\langle \sigma v\rangle_{\chi\chi\rightarrow \gamma\,\gamma} \approx 1.27\times10^{-27}{\rm cm^3/s}$ for the Einasto profile (normalized to $\rho_{dm}^{sun}=0.4\,{\rm GeV/cm^3}$) \cite{1204.2797}. 
After a rescaling as in \cite{1101.2610} this corresponds to a $Z\gamma$ annihilation cross-section of \ba \label{reqd-Zgamma}\langle \sigma v\rangle^{{\rm reqd}}_{\chi\chi\rightarrow Z\,\gamma} \approx 3.1\times10^{-27}{\rm cm^3/s} \;\;-{\rm Einasto}. \ea Note that for simplicity here we have neglected the contribution of the $\gamma\,\gamma$ channel to the signal, since it is suppressed relative to $Z\,\gamma$. Including this channel will change the above values slightly. 

An important point worth remembering is that this effective cross-section \emph{assumes} that the entire DM abundance consists of WIMPs. 
However, from a string/$M$ theory perspective one expects that both WIMPs and axions form a significant fraction of DM. Using the information about the WIMP mass from the $\gamma$-line signal, we can constrain this fraction (up to the uncertainties explained in section \ref{implications} and below). To set up the notation, we define: \ba \label{frac}\Omega_{\chi}=\eta\,\Omega_{dm};\;\;\;J_{\chi}=\eta_{GC}^2\,J_{dm},\ea where $J_A\equiv\,\int_{l.o.s,\,\Delta\Omega}ds\,d\Omega\,\rho_A^{2}(r)$. $\eta$ is the fraction of the total DM abundance in the form of WIMPs, and $\eta_{GC}$ is the fraction in WIMPs along the line of sight (toward the GC). $\eta$ does not equal $\eta_{GC}$ in general;
for example, the DM fraction in the form of WIMPs along the line-of-sight may itself depend non-trivially on the galactocentric distance $r$ due to the existence of ``boost"(clump) factors. Also, the uncertainty in the DM density at the position of the sun can be folded into $\eta_{GC}$ if one normalizes to the value $0.4\,{\rm GeV/cm^3}$ chosen in \cite{1204.2797}. 

The annihilation cross-section of a  light Wino LSP to $Z\,\gamma$ is a loop process with charged Winos (almost degenerate with the LSP ) and $W$-bosons in the loop. Hence the magnitude of the cross-section is determined by the LSP mass $m_{\chi}$ and the $SU(2)$ gauge coupling $g$, and was first computed in  \cite{Ullio:1997ke}.  For a Wino mass of 145 GeV, we estimated the cross-section using DarkSUSY \cite{Gondolo:2004sc}, which uses the expressions in \cite{Ullio:1997ke}: \ba\label{annih} \langle \sigma v\rangle_{\chi\chi\rightarrow Z\,\gamma} \approx 1.26\times 10^{-26}\,{\rm cm^3/s}\ea 
The total annihilation cross-section of Wino-like LSPs is dominated by the annihilation to $W^+ \,W^-$; however due to the presence of almost degenerate charginos split only by around 160 MeV \cite{Gherghetta:1999sw}, coannihilation effects are also important in the early Universe (although not at present). Contrary to the thermal case where the coannihilation contribution is determined by the freezeout temperature $T_F$, in this case the coannihilation is determined by the reheat temperature $T_{R}$. We have computed the total cross-section using both DarkSUSY\cite{Gondolo:2004sc} and MicrOMEGAs \cite{Belanger:2006is}. 
\begin{figure}[t!]
\includegraphics[width=3.45in]{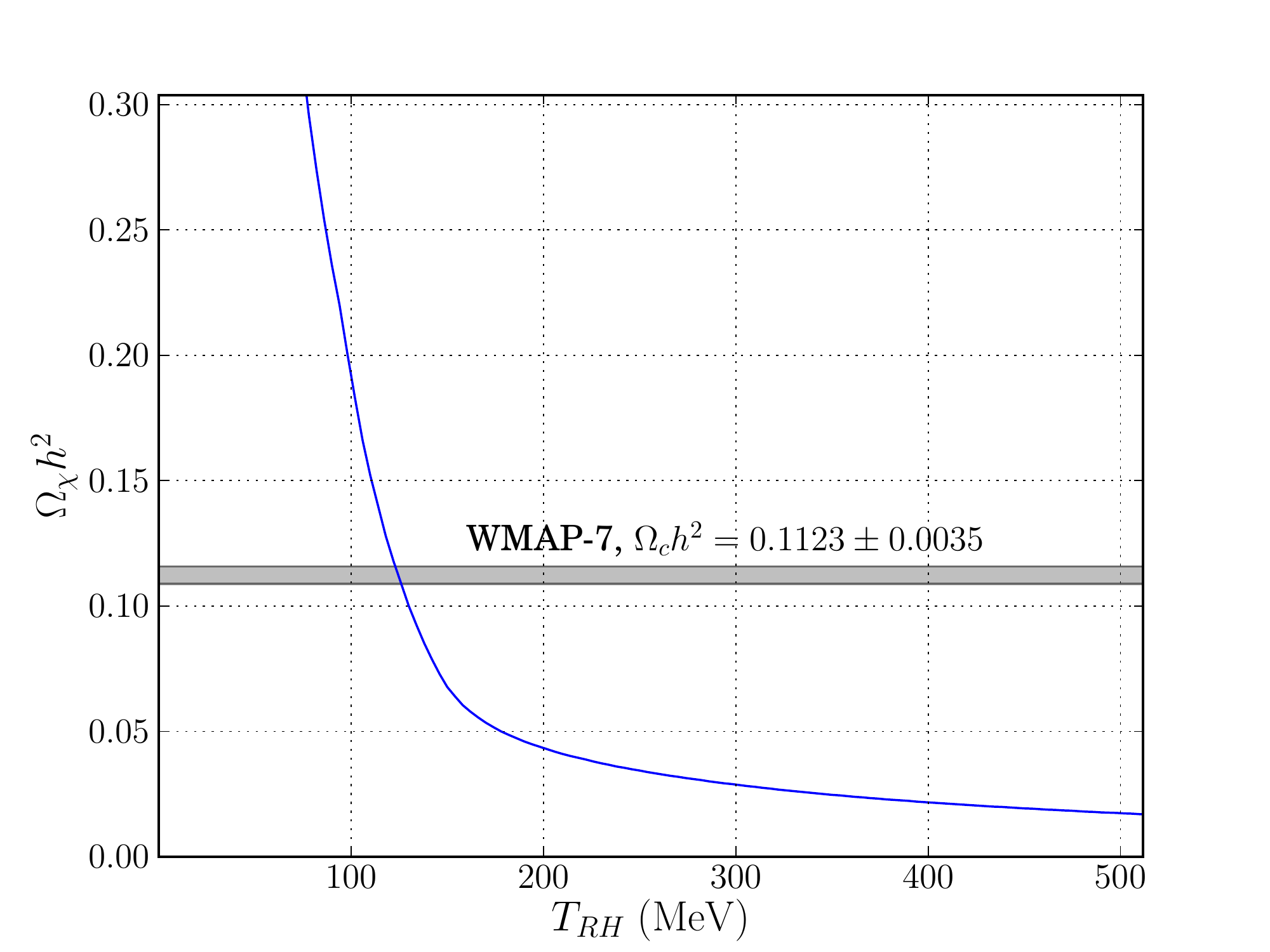}
\caption{\footnotesize{Relic Abundance within a non-thermal cosmological history of a Wino-like LSP with mass 145 GeV as a function of the reheat temperature $T_R$ arising from the decay of the lightest modulus.}}\label{relic}
\end{figure}

Since the $\gamma$-ray signal is proportional to $J_{\chi}\,\langle \sigma\,v\rangle_{\chi\chi\rightarrow Z\gamma}$ from (\ref{flux}), using (\ref{reqd-Zgamma}), (\ref{frac}), and (\ref{annih}) one gets: \ba\label{eta-GC} \eta_{GC} \approx \sqrt{\frac{3.1\times10^{-27}}{1.26\times 10^{-26}}} \approx 0.49^{+0.07}_{-0.08}\;\;{\rm -\,Einasto}\ea where the uncertainties come from those estimated in \cite{1204.2797}. 
We will use the Einasto profile as a benchmark as this provides more stringent constraints compared to the NFW or isothermal profiles, as will be seen in section \ref{associated}.

Since $\eta_{GC} < 1$, when interpreted within this framework, the signal implies that DM does not consist only of WIMPs. {\it Furthermore, if $\eta \approx \eta_{GC}$, then for an Einasto profile approximately $49\%$ of the DM consists of WIMPs and the other $51\%$ consists of axions}. However, these fractions will change if $\eta \neq \eta_{GC}$. Note that a very promising proposal has been made for measuring the (ultra-light) axion fraction of dark matter using weak lensing tomography \cite{Marsh:2011bf} and the results of this are keenly anticipated, see also \cite{Marsh:2012nm}.

One can also calculate theoretically the abundance of the Wino-like LSPs $\Omega_{\chi}$ and that of the axions $\Omega_{a}$ within a non-thermal cosmological history. Starting with WIMPs, from the arguments leading to (\ref{ntwm}), it can be shown that \cite{Acharya:2008bk}: \ba \label{wimp}\Omega_{\chi}\,h^2 \approx \frac{45}{2\pi \sqrt{10\,g_{\star}(T_R)}}\,\frac{m_{\chi}}{(\rho/s)_{crit}}\,\left(\frac{1}{m_{pl}\,T_R\,\langle \sigma\,v\rangle_{\chi\chi}^{tot}}\right)\ea where $m_{pl}$ is the reduced Planck scale, $g_{\star}(T_R)$ measures the relativistic degrees of freedom at $T_R$, and as explained above $\langle \sigma\,v\rangle_{\chi\chi}^{tot}$ is also a function of $T_R$ which is computed using MicrOMEGAs. Then, the LSP abundance $\Omega_{\chi}$ can be computed as a function of $T_R$ as in Figure \ref{relic}. From the Figure, we see that for $100\,{\rm MeV} \lesssim T_R \lesssim 300$ MeV, the relic abundance changes quite rapidly due to the rapidly changing value of $g_{\star}(T_R)$ caused by the QCD phase transition.

The relic abundance for the axions on the other hand is given by \cite{Acharya:2010zx}: \ba\label{axions} \Omega_a\,h^2 \approx {\cal O}(1)\,\frac{T_R}{(\rho/s)_{crit}}\,\left(\frac{f_a}{m_{pl}}\right)^2\,\langle \theta^2\rangle,\ea where $f_a$ is the axion decay constant and $\theta$ is the axion misalignment angle in the early Universe. Then requiring that $\Omega_{\chi}=\eta\,\Omega_{dm}$ and $\Omega_{a}=(1-\eta)\,\Omega_{dm}$, one gets from (\ref{axions}) : 
\ba \label{TR} T_R \approx \frac{4 \times 10^{-6}{\rm GeV}\,(1-\eta)}{\langle \theta^2\rangle}\ea  where we have used $(\rho/s)_{crit}=3.6\times 10^{-9}\,{\rm GeV}$, and have taken $f_a \approx 10^{-2}\,m_{pl} \sim M_{GUT}$ as is natural in realistic string compactifications \cite{Acharya:2010zx}.

Thus for $\eta \approx \eta_{GC}$, from Figure \ref{relic} and eqn. (\ref{TR}), one obtains: \ba
T_R &\approx& 166 \,{\rm MeV}\,\;\;{\rm with}\,g_{\star}(T_R)\approx 74 \nonumber\\ \langle\theta^2\rangle^{1/2} &\approx& 4.\times 10^{-3}\ea which corresponds to $\langle \sigma\,v\rangle_{\chi\chi}^{tot} \approx 5.6\times10^{-24}\,{\rm cm^3/s}$.  The misalignment angle is thus tuned between the per cent and per mille  level, which is much better than that in the thermal case for many axions. Again, if $\eta \neq \eta_{GC}$ these numbers will change, though the qualitative result will likely remain unchanged. Since $T_R = (\frac{90}{\pi^2\,g_{\star}(T_R)})^{1/4}\left(\frac{m_X^3\,m_{pl}}{f_{X}^2}\right)^{1/2}$, where $f_{X}$ is the decay constant of the lightest modulus $X$ (which has a mass $m_{X}\approx 2\,m_{3/2}$ \cite{1204.2795, Acharya:2008bk}), this means that for example $f_{X} \approx 2.3\times 10^{17}$ GeV for $m_{3/2}=60$ TeV, which is close to the string or eleven dimensional scale.

\section{Constraints and Associated WIMP Signals}\label{associated}

\begin{figure}[h!]
\includegraphics[width=3.45in]{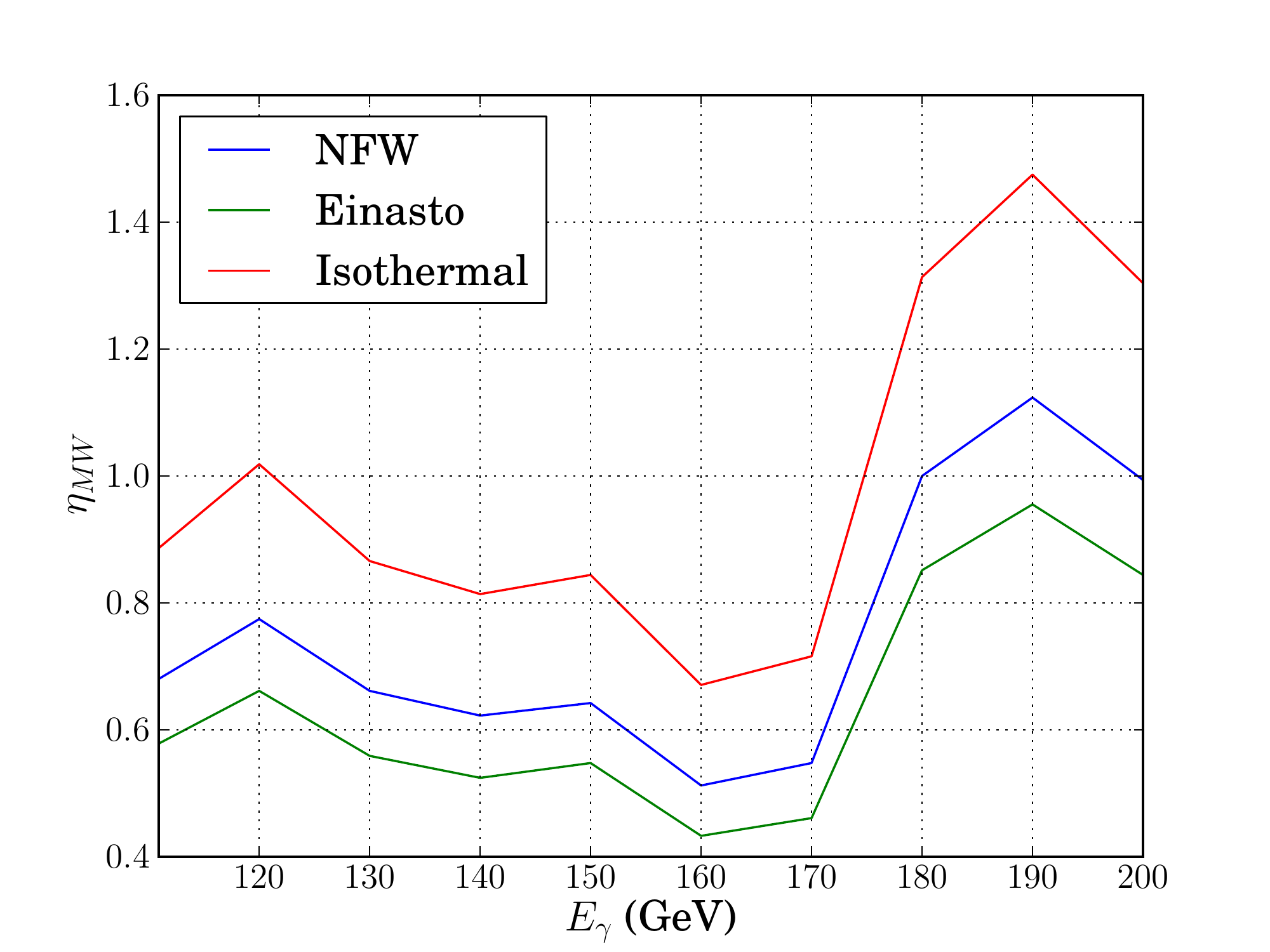}
\caption{\footnotesize{Upper bound on $\eta_{MW}$, defined in a similar way as in (\ref{frac}) for the Milky Way, as a function of the $\gamma$-ray energy for WIMPs $\chi$ annihilating to $Z\gamma$. $\eta_{MW}$ provides a rough estimate of the \emph{overall} WIMP fraction of DM, as explained below.}}\label{eta-MW}
\end{figure}

As mentioned in section \ref{signal}, the Fermi-LAT collaboration has put strong upper limits on the annihilation cross-section to $Z\gamma$ by searching for $\gamma$-ray lines from essentially the entire Milky Way \cite{Ackermann:2012qk}. Within our framework, WIMPs are only part of the total DM. Hence, the limits roughly put an upper bound on the fraction of DM in the form of WIMPs. More precisely, defining $J_{\chi}^{MW}=\eta_{MW}^2\,J_{dm}$ in a similar way as in (\ref{frac}) but now for the Milky Way, the limits in \cite{Ackermann:2012qk} put an upper bound on $\eta_{MW}$ as seen in Figure \ref{eta-MW}, 
Note that since $\eta\neq \eta_{MW}$ in general, technically this does not put an upper bound on the \emph{overall} LSP fraction of DM, but it is still expected to provide a reasonable estimate. Thus, the fact that $\eta_{MW} < 1$ for \emph{all} three profiles for $E_{\gamma}\approx 130$ GeV ($m_{\chi}\approx 145$ GeV) is remarkably consistent with an additional
component of DM.

If $\eta\approx \eta_{GC}\approx \eta_{MW}$, then for $m_{\chi}=145$ GeV ($E_{\gamma}=130$ GeV), this implies an upper bound on the LSP fraction of DM $\eta_{max} (m_{\chi}=145\,{\rm GeV}) \approx 0.56$ for the Einasto profile, which is consistent with the fraction obtained ($0.49$) assuming the existence of the $\gamma$-line signal  (see statement below (\ref{eta-GC})).
The fact that the Fermi-LAT limits are comparable to the results we obtain implies that their analysis is quite close to being sensitive to the Wino signal. Note that the constraints from the NFW and isothermal profiles are \emph{weaker} than those for the Einasto profile, so  they are also easily compatible with the $\gamma$-line signal. Finally, for $E_{\gamma} \gtrsim 180$ GeV ($m_{\chi} \gtrsim 191$ GeV) for the NFW profile and for $E_{\gamma} \gtrsim 175$ GeV for the isothermal profile, $\eta_{MW}$ becomes larger than unity, implying that there is no bound.

The non-observation of excess \emph{diffuse} $\gamma$-rays from the GC and dwarf spheroidal galaxies (dSphs) also puts constraints on the framework in general. The bounds from the GC can be easily satsified, however the bounds from dwarf spheroidals provide non-trivial constraints. In particular, Fermi-LAT has put bounds on the total cross-section of WIMPs, assuming a dominant annihilation to $W$ pairs \emph{and} assuming that WIMPs form the entire DM content of the Universe. For $m_{\chi}=145$ GeV, the bound is $1\times 10^{-25}\,{\rm cm^3/s}$ \cite{Garde:2011hd}.  Again, since we expect that WIMPs are only part of the total DM and since the total cross-section \emph{at present} is smaller than that in the early Universe because coannihilation effects are absent now (the temperature inside the dwarf galaxies is smaller than the mass splitting between the LSP and lightest chargino), the bounds on the total WIMP cross-section are relaxed.  Defining $J_{\chi}^{dSph}=\eta_{dSph}\,J_{dm}$ in a similar way as in (\ref{frac}) but now along the line of sight for the dwarf galaxies, one finds that the bound on the total WIMP annihilation cross-section becomes:\ba \langle \sigma\,v\rangle^{max}\approx \frac{1\times10^{-25}\,{\rm cm^3/s}}{\eta_{dSph}^2}.\ea For $\eta_{dSph}\approx \eta_{GC}$, this gives $2.6\times 10^{-25}\,{\rm cm^3/s}$, which is in tension with the value obtained for a Wino-like LSP -- $3\times 10^{-24}\,{\rm cm^3/s}$ \cite{Gondolo:2004sc} (since coannihilation effects are not important at present). However, this tension can be relaxed if $\eta_{dSph} < \eta_{GC}$, say $\eta_{dSph} \lesssim 0.37\,\eta_{GC}$. Presumably, this also implies that one expects to see a diffuse  $\gamma$-ray DM signal from dwarf spheroidal galaxies in the near future; conversely, non-observation would place 
severe constraints on the framework.

The Wino-like LSP interpretation of the tentative $\gamma$-line signal has consequences for WIMP signals in other astroparticle-physics observables. First, as mentioned in section \ref{signal-string}, Winos also give rise to a $\gamma\,\gamma$ signal which will give rise to a (smaller) peak at $E_{\gamma} \approx 145$ GeV. This should be eventually seen by Fermi and AMS2. The $\gamma$-line signal should also be observable from dwarf spheroidal galaxies with sufficient data. In addition, for a given $\eta$, it is possible in principle to make a prediction for the cosmic ray positron fraction measured by PAMELA \cite{Adriani:2008zr} and Fermi \cite{FermiLAT:2011ab}, as well as the total electron and positron signal measured by Fermi \cite{Ackermann:2010ij}. However, this is complicated in practice for the following reason. Given the Wino interpretation, the positron fraction as well as the total electron and positron flux cannot keep on rising above 145 GeV. So, although it may be possible to explain the positron fraction measured by PAMELA (possibly with a mild boost factor), the fact that Fermi sees both a large total electron and positron flux up to around 800 GeV, as well as a rising positron fraction up to about 200 GeV implies that there must be an additional source of electrons and positrons. This (unknown) source will in general also modify the prediction for the positron fraction between 10 and 100 GeV (as measured by PAMELA), so it is not possible to make a reliable statement without assuming a magnitude and shape of the additional unknown contribution. Hence we leave this issue aside for future investigation.

\section{Gluino and Bino Mass Prediction}\label{falsifiable}

Theoretically,  a Wino-like LSP with a mass of 145 GeV strongly constrains other parameters (such as the Gluino mass), since in the string/$M$ theory framework all of these parameters are determined by a few ``microscopic" quantities which are determined by the extra dimensions (see below). 
Therefore, this leads to correlated predictions for other observables as well. In particular, the subject of this section is  the prediction for the Gluino and Bino mass consistent with a  Wino-like LSP whose mass is 145 GeV. The framework summarized in section \ref{string} attempts to describe a large class of  string/$M$ theory vacua. 
For concrete gaugino mass calculations, we will here focus on the $M$ theory vacua studied in \cite{arXiv:0801.0478, Acharya:2006ia, Acharya:2008hi}. These vacua naturally give rise to Grand Unified Theories and 
for concreteness, we will assume the matter and gauge spectrum below the GUT scale is that of the MSSM, so that our precise results are valid for the \emph{same} framework which gives a prediction for the Higgs mass as in \cite{1204.2795}.

In $M$ theory compactifications without fluxes, the entire scalar potential is generated through non-perturbative effects 
(arising from strong dynamics in the hidden sector) and depends on all the moduli. This potential stabilises all the moduli and 
generates an exponential hierarchy between the Planck scale and the gravitino mass $m_{3/2}$ i.e. the scale of supersymmetry breaking. 
One obtains \cite{1204.2795}: 
\ba \alpha_h^{-1} &=& {1 \over 2\pi} {PQ \over Q-P}\,\log\left({A_1 Q \over A_2 P}\right);\\ \alpha_{vis}^{-1} &=& \sum_i\,N_i^{vis}\,s_i ;\;s_i \sim {1 \over N_i} {1 \over \alpha_h};\;i=1,.,N;\,\nonumber\\ \frac{m_{3/2}}{m_{pl}} &=& e^{K/2}\,\frac{W}{m_{pl}^3}=A_2 {|Q-P| \over Q} \alpha_h^{7/2} e^{-\frac{2\pi}{Q \alpha_h}}\nonumber
\ea Here $s_i$ are the moduli vevs, $P$ and $Q$ are the ranks of the hidden sector gauge groups, $A_1,A_2$ are ${\cal O}(1)$ constants, and $N_i,N_i^{vis}$ are positive integers. Thus, all the moduli \emph{vevs}, the hidden sector and visible sector gauge couplings $\{\alpha_h, \alpha_{vis}\}$, and gravitino mass $m_{3/2}$ are determined in terms of the dimensionless microscopic quantities $\{P,Q,A_1,A_2,N_i\}$. Note that $m_{pl}$ is the only dimensionful parameter in these formulae.

As reviewed in \cite{1204.2795}, supersymmetry breaking is dominated by a field which is {\it not} a geometric modulus. 
This implies that the gaugino masses, which arise from the $F$-terms of geometric moduli, do not get contributions from the dominant $F$-term, implying that they are suppressed relative to $m_{3/2}$. The tree-level gaugino masses at $M_{GUT}$ are universal (since the SM arises from a GUT) and can be determined in terms of the microscopic quantities as an expansion in the small (asymptotically free) hidden gauge coupling $\alpha_h$ \cite{Acharya:2008hi}:\ba\label{gaugino} M_{1/2}^{tree}(M_{GUT})&=& \sum_{i=1}^{N}\,\frac{F^i\,\partial_i\,f_{vis}}{2\,i\,{\rm Im}(f_{vis})}\simeq-\frac{\alpha_h\,Q}{6\,\pi}\,m_{3/2}\left(1+...\right) \nonumber\\
&\simeq& -\frac{1.9\,m_{3/2}}{P_{eff}}\left(1+\epsilon\right)\ea Here,  $P_{eff}\equiv P\,\log\left(\frac{A_1Q}{A_2P}\right)$ and ${\rm Im}(f_{vis}) \equiv \alpha_{vis}^{-1} = \sum_i\,N_i^{vis}\,s_i$. Furthermore, it can be shown that the suppression factor $P_{eff}$ in (\ref{gaugino}) is determined once the requirement of approximate vanishing of the cosmological constant is imposed! In particular, this results in $P_{eff}\approx 61.65$ \cite{Acharya:2008hi}. Thus the tree-level gaugino mass is completely determined at the leading order! 

The subleading correction $\epsilon$ in (\ref{gaugino}) is a combination of a number of factors: the threshold correction to $\alpha_{GUT}$ (to which the contributions from Kaluza-Klein modes have been computed in \cite{Friedmann:2002ty}), 
the higher order corrections proportional to $\alpha_h$ and computed in \cite{Acharya:2008hi}, and the higher order corrections in the K\"{a}hler potential for the matter field with the dominant $F$-term which depend on more details of the compactification and are harder to compute. In addition, there could be threshold corrections to the gauge coupling from four-dimensional states near the GUT scale such as the Higgs triplets arising in an $SU(5)$ GUT.  The computable contributions to $\epsilon$ generically give it a value with magnitude smaller than unity; hence, here we treat $\epsilon$
as a free parameter with magnitude as such. 

Furthermore, since the tree-level gaugino masses are suppressed relative to $m_{3/2}$, the anomaly mediated contributions are also important. These do not depend on any new microscopic quantities, but make the gaugino masses non-universal. Thus, the gaugino masses at $M_{GUT}$ depend only on the quantities -$\{m_{3/2},\epsilon\}$, where $m_{3/2}$ varies from $\sim 30$ TeV to $\sim 80$ TeV depending on the microscopic quantities, and $\epsilon$, which is a number with magnitude smaller than unity \cite{Acharya:2008hi}.  
{\renewcommand{\arraystretch}{1.5}
\begin{table}[h!]
\begin{tabular}{ | p{3.8 cm} |p{4.1cm}|}
\hline
$30\,{\rm TeV}\leq m_{3/2}\leq 80\,{\rm TeV}$ &  $0.05\,m_{3/2}\leq |\mu| \leq 0.15\,m_{3/2}$\\
$-0.6 \leq \epsilon \leq 0.2$&  $1.0\,m_{3/2} \leq A_t \leq 1.5\,m_{3/2}$ \\
\hline
\end{tabular}
\caption{\footnotesize{Variation of Theoretical Inputs in ranges consistent with theoretical expectation. Only $\mu$ suppressed relative to $m_{3/2}$ (``small $\mu$") is considered since unsuppressed $\mu$ gives rise to a Higgs mass which is ruled out.  $\tan\beta$ is correlated with $\mu$ via EWSB; $\mu$ in the above range gives $7 \lesssim \tan \beta \lesssim 15$. For more details on choices of these quantities, see \cite{1112.1059}.}}\label{inputs} 
\end{table}}

\begin{figure}[t!]
\includegraphics[width=3.45in]{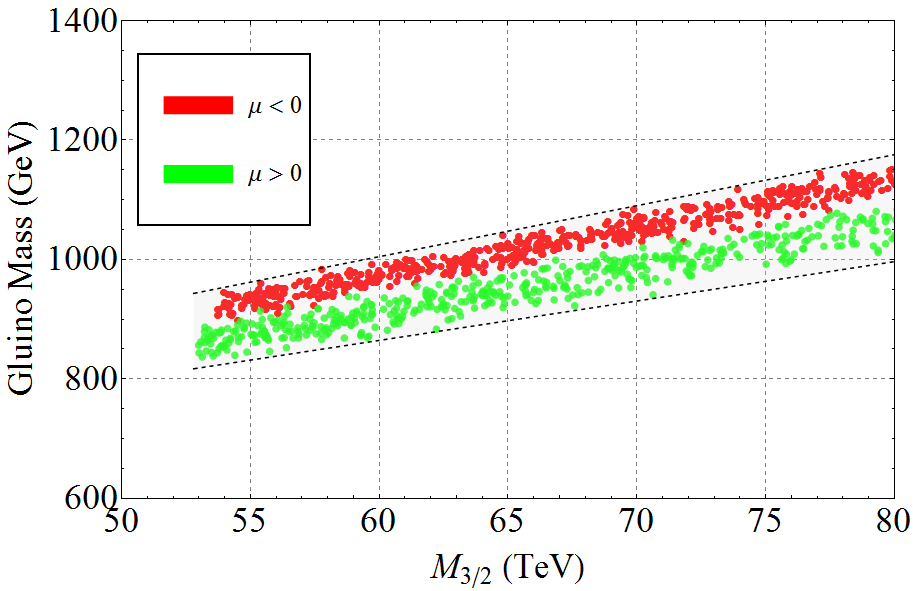}
\caption{\footnotesize{Gluino Pole Mass computed with SPheno as a function of $m_{3/2}$ for ${\rm sign}(\mu) > 0$ (green) and ${\rm sign}(\mu) < 0$ (red) with Wino-like LSPs in the range $140\,{\rm GeV}\lesssim m_{\chi} \lesssim 150$ GeV, and with other inputs varied in ranges specified in Table \ref{inputs}.}}\label{M3-m32}.
\end{figure}

\begin{figure}[t!]
\includegraphics[width=3.45in]{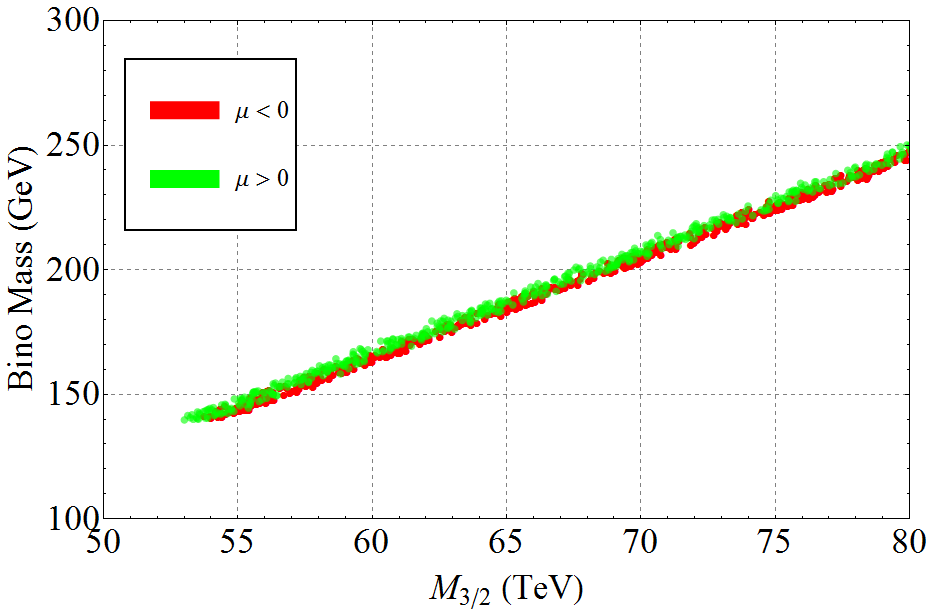}
\caption{\footnotesize{Bino Pole Mass computed with SPheno as a function of $m_{3/2}$ for ${\rm sign}(\mu) > 0$ (green) and ${\rm sign}(\mu) < 0$ (red) with Wino-like LSPs in the range $140\,{\rm GeV}\lesssim m_{\chi} \lesssim 150$ GeV, and with other inputs varied in ranges specified in Table \ref{inputs}.}}\label{M1-m32}
\end{figure}

The gaugino pole masses, however, depend on other quantities due to renormalization group evolution (RGE). 
These are the $\mu$ parameter, the trilinear coupling for the third generation ($A_t$) and the ratio of the Higgs {\it vevs} ($\tan\beta$), all of which are also determined in terms of the microscopic quantities after moduli stabilization. 
The dependence on the $\mu$ parameter arises from the Higgsino leading-log threshold, while that on $A_t$ arises from the dependence of the RGE for the gaugino masses on $A_t$ at two-loops. The Higgsino threshold also depends on the sign of $\mu$, so the gaugino pole masses depend on ${\rm sign}(\mu)$ as seen in Figures \ref{M3-m32} and \ref{M1-m32}. $\tan\beta$, $\mu $ and $A_t$ are correlated via electroweak symmetry breaking (EWSB). In particular, $A_t={\cal O}(1)\,m_{3/2}$ and  $\mu$ suppressed relative to $m_{3/2}$ (``small $\mu$") gives rise to $\tan\beta \gtrsim 5$, while unsuppressed $\mu$ gives $\tan\beta \lesssim 5$, as described in \cite{1112.1059}.  The variation of these quantities is described in Table \ref{inputs}; for more details on this, and the procedure for the two-loop RG evolution of the soft parameters from the GUT scale to the electroweak scale and including the relevant threshold corrections to compute the pole masses, see \cite{1112.1059} and references therein. Since $m_{3/2}\gtrsim 30$ TeV, with $\mu$ in the range as  in Table I, the LSP (and the second lightest neutralino) have very small Higgsino components and are predominantly gaugino like. For spectrum and EWSB calculations, we used the numerical codes SoftSUSY \cite{Allanach:2001kg}and SPheno \cite{Porod:2003um} finding good agreement between the two.

Figures \ref{M3-m32} and \ref{M1-m32} show the predictions for the Gluino mass $m_{\tilde{g}}$ and Bino mass $m_{\tilde{b}}$ as functions of $m_{3/2}$ when the inputs are varied as in Table \ref{inputs}, for choices of microscopic quantities which yield a Wino-like LSP mass in the range $140\;{\rm GeV} \lesssim m_{\chi} \lesssim 150$ GeV. The green (red) scatter points correspond to models with $\mu > 0\,(< 0)$ respectively. We find $900\,{\rm GeV}\lesssim m_{\tilde{g}} \lesssim 1180$ GeV for $\mu < 0$ when $m_{3/2} \lesssim 80$ TeV, while $820\,{\rm GeV}\lesssim m_{\tilde{g}} \lesssim 1100$ GeV for $\mu > 0$ for the same range of $m_{3/2}$. The corresponding prediction for the Bino mass is: $145\,{\rm GeV}\lesssim m_{\tilde{b}} \lesssim 250$ GeV for either sign of $\mu$. Note that for a given $m_{3/2}$, the spread in the Gluino mass arising from the variation of the other quantities in Table \ref{inputs} is $\lesssim 50$ GeV for negative $\mu$ and slightly larger for positive $\mu$. For Binos, the spread is much smaller:  $\lesssim 10$ GeV for both signs of $\mu$. The upper limits of the gaugino masses may be extended a bit if $m_{3/2}$ is slightly larger than 80 TeV. However, we do not expect it to be much larger because then the Higgs mass will be too large \cite{1112.1059} and the fine-tuning required to keep the axion abundance consistent with observations will also increase \cite{Acharya:2010zx,1204.2795}. 

\begin{figure}[h!]
\includegraphics[width=3.45in]{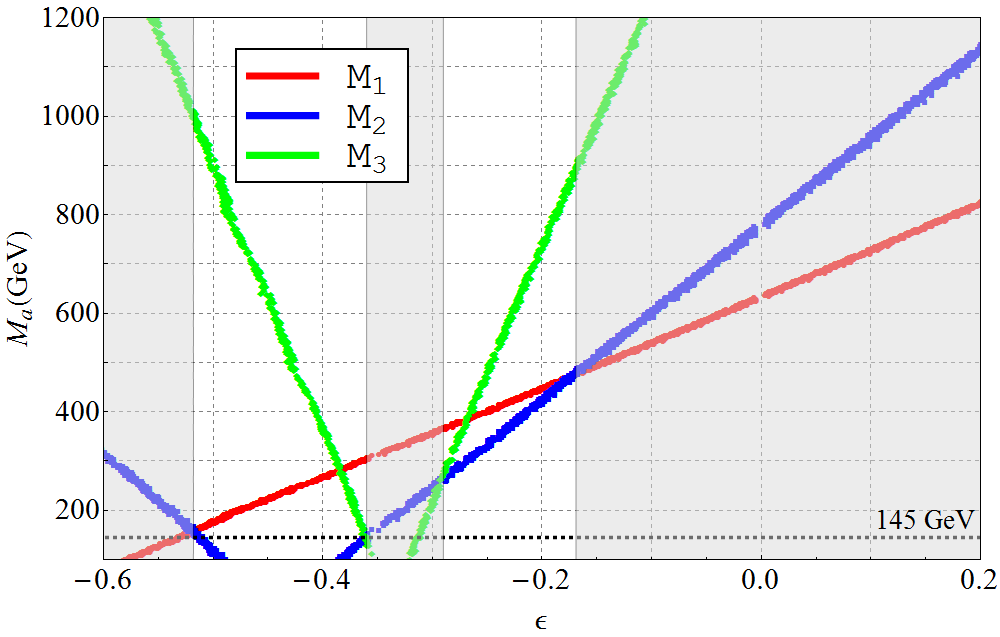}
\caption{\footnotesize{Bino, Wino and Gluino Pole Masses computed with SPHENO as a function of $\epsilon$ for $m_{3/2} = 60$ TeV and ${\rm sign}(\mu) < 0$, and with other inputs varied in the ranges specified in Table \ref{inputs}. The shaded region corresponds to cases in which the LSP is not Wino-like.}}\label{Ma}
\end{figure}
It is also useful to study Figure \ref{Ma}, which shows the result for the three gaugino pole masses as a function of the parameter $\epsilon$ in (\ref{gaugino}) for $m_{3/2}=60$ TeV, ${\rm sign}(\mu) < 0$, and varying the other inputs in the ranges specified in Table \ref{inputs}. In particular, the blue points which correspond to Wino LSPs, arise for $-0.52 \lesssim \epsilon \lesssim -0.36$, and for  $-0.28 \lesssim \epsilon \lesssim -0.17$. However, a 145 GeV  Wino-like LSP with $m_{3/2}=60$ TeV only exists when $\epsilon \simeq -0.52$ or $-0.36$ with the latter being excluded by a Gluino which is too light.

\section{Conclusion}
In conclusion, taking seriously the tentative $\gamma$-line signal in \cite{1204.2797}, we have shown that - consistent with the generic predictions in \cite{1204.2795} - it is possible to naturally explain the signal as due to the annihilation of Wino-like LSPs which constitute roughly half of the total DM content and which is compatible with all current constraints. The string/$M$ theory framework naturally predicts that the remaining fraction of DM is in the form of axions. 
Finally, correlated falsifiable predictions for astrophysics and particle physics observables can also be made. Further investigations of the results reported in \cite{1204.2797} are eagerly anticipated.

From a theoretical point of view, clearly more needs to be understood. We have found that  in cases with sufficiently conserved R-parity (or any analagous symmetry), the string/$M$ theory vacua which are phenomenologically viable are those in which gaugino masses are
suppressed. Furthermore, within this class, only vacua with Wino-like LSPs are viable,
in that Wino-like LSPs form a significant fraction of DM, consistent with all constraints and can explain the tentative Fermi-LAT signal. 
That such features arise in a reasonably large region of the microscopic parameter space is encouraging, and
it is very important that we improve our understanding of the microscopic aspects of the theory.

\acknowledgements{P.K. thanks K. Bobkov, and B.S.A. thanks A. Hryczuk, R. Iengo, and P. Ullio for helpful conversations. B.S.A, G.K, P.K, R.L, and B.Z would like to thank the Simons Center for Geometry and Physics for hospitality during part of the work. The work of G.K, R.L, and B.Z is supported by the DoE Grant DE-FG-02-95ER40899 and by the MCTP. R.L. is also supported by the String Vacuum Project Grant funded through NSF 
grant PHY/0917807. The work of PK is supported by the DoE Grant DE-FG02-92ER40699. B.S.A. gratefully acknowledges the support of the Science and Technology Facilities Council.}

\end{document}